\documentclass[prx,a4paper,aps,twocolumn,nofootinbib,superscriptaddress,10pt,showkeys]{revtex4-1}

\usepackage{amsmath,amsthm,amsfonts,graphicx,xcolor,times,xfrac,booktabs,mathtools,enumitem,xr,subfigure,amssymb,bbm,verbatim,appendix,placeins}
\usepackage[unicode=true,bookmarks=true,bookmarksnumbered=false,bookmarksopen=false,breaklinks=false,pdfborder={0 0 1}, backref=false,colorlinks=true]{hyperref}
\usepackage{orcidlink}
\usepackage{bm}

\usepackage{graphicx}
\renewcommand*{\url}[1]{\href{#1}{#1}}

\makeatletter
\theoremstyle{plain}

\theoremstyle{plain}

\theoremstyle{plain}

\theoremstyle{remark}
\newtheorem*{rem*}{\protect\remarkname}
\theoremstyle{plain}

\theoremstyle{plain}

\theoremstyle{definition}
\newtheorem{defn}{\protect\definitionname}
\theoremstyle{plain}

\theoremstyle{plain}
\newtheorem*{thm*}{\protect\theoremname}
\theoremstyle{plain}
\newtheorem*{lem*}{\protect\lemmaname}
 
\providecommand{\propositionname}{Proposition}
\providecommand{\theoremname}{Theorem}
\providecommand{\lemmaname}{Lemma}
\providecommand{\remarkname}{Remark}
\providecommand{\conjecturename}{Conjecture}
\providecommand{\definitionname}{Definition}
\providecommand{\corollaryname}{Corollary}
\providecommand{\observationname}{Observation}
\allowdisplaybreaks

\def\ket#1{\vert{#1}\rangle}

\def\BraVert{e.g.,roup\,\mid\,\bgroup}

\def\ketbra#1#2{\vert{#1}\rangle\!\langle{#2}\vert}

\def\tr#1{\operatorname{tr}\left[{#1}\right]}
\newcommand{\ptr}[2]{\operatorname{tr}_{#1}\left[ #2 \right]}

\newcommand{\Ecal}{\mathcal{E}}

\newcommand{\Lcal}{\mathcal{L}}

\newcommand{\Scal}{\mathcal{S}}

\newcommand{\Dcal}{\mathcal{D}}

\newcommand{\id}{\mathbbm{1}}

\let\oldaddcontentsline\addcontentsline

\newcommand{\starttocentries}{\let\addcontentsline\oldaddcontentsline}

\newcommand{\cpt}{\mathcal{E}}

\newcommand{\dynamics}[1]{(#1)}

\newcommand{\Sys}{\mathcal{S}} 
\newcommand{\Anc}{\mathcal{A}} 
\newcommand{\Mem}{\mathcal{M}} 

\newcommand{\Dyn}{\mathcal{D}}

\newcommand{\SA}{{\Sys\!\Anc}}
\newcommand{\rhoSA}{\rho^\SA}
\newcommand{\rhoSAzero}{\rho^\SA_0}
\newcommand{\rhoSAone}{\rho^\SA_{t_1}}
\newcommand{\rhoSAtwo}{\rho^\SA_{t_2}}
\newcommand{\sigmaSA}{\bm{\sigma}}

\newcommand{\op}[1]{\hat{#1}}
\newcommand{\covmat}{\bm{\sigma}}

\def\Hilb{\ensuremath{\mathcal{H}}}
\def\e{\ensuremath{\mathrm{e}}}
\def\iu{\ensuremath{\mathrm{i}}}
\def\i{\iu}

\def\Hilb{\ensuremath{\mathcal{H}}}

\newcommand{\comm}[1]{\left[#1\right]}
\newcommand{\abs}[1]{\left|#1\right|}

\renewcommand{\d}{\mathrm{d}}

\theoremstyle{definition}
\newtheorem{theorem}{Theorem}

\begin{document}

\author{Charlotte Bäcker}
\affiliation{Institute of Theoretical Physics, TUD Dresden University of Technology, 01062, Dresden, Germany}
\author{Konstantin Beyer}
\affiliation{Institute of Theoretical Physics, TUD Dresden University of Technology, 01062, Dresden, Germany}
\affiliation{Department of Physics, Stevens Institute of Technology, Hoboken, New Jersey 07030, USA}
\author{Walter T. Strunz}
\affiliation{Institute of Theoretical Physics, TUD Dresden University of Technology, 01062, Dresden, Germany}

\title{Entropic witness for quantum memory in open system dynamics}
\date{\today}

\date{\today}
\begin{abstract}
The dynamics of open quantum system are often modeled by non-Markovian processes that account for memory effects arising from interactions with the environment. It is well-known that the memory provided by the environment can be classical or quantum in nature.
Remarkably, the quantumness of the memory can be witnessed locally by measurements on the open system alone, without requiring access to the environment. 
However, existing witnesses are computationally challenging for systems beyond qubits. 
In this work, we present a tractable criterion for quantum memory based on the von Neumann entropy, which is easily computable for systems of any dimension. 
Using this witness, we investigate the nature of memory in a class of physically motivated finite-dimensional qudit dynamics.
Moreover, we demonstrate that this criterion is also suitable for detecting quantum memory in continuous-variable systems. As an illustrative example, we analyze non-Markovian Gaussian dynamics of a damped harmonic oscillator.
\end{abstract}


\maketitle

\section{Introduction}

Non-Markovian dynamics in open quantum systems have attracted significant attention in recent decades.
While many open quantum systems are adequately described by Markovian dynamics, particularly through the Gorini-Kossakowski-Sudarshan-Lindblad (GKSL) master equation, numerous theoretical and experimental advances have underscored the need to study the rich quantum dynamics that go beyond these frameworks~\cite{liuExperimentalControlTransition2011, SmiMegVac2021, RivHuePle2014, LiGuoPii2019, ApoDiPlaPat2011, HaiMcEDePalMan2011, VasOliParMan2011, PolRodFraPatMod2018,thorwartEnhancedQuantumEntanglement2009,strathearnEfficientNonMarkovianQuantum2018,groblacherObservationNonMarkovianMicromechanical2015,hartmannExactOpenQuantum2017,tanimuraNumericallyExactApproach2020,boettcherDynamicsStronglyCoupled2024,koyanagiClassicalQuantumThermodynamics2024,linkOpenQuantumSystem2024}. 
Non-Markovian behavior always relies on some sort of memory effect. 
In the typical open quantum systems picture, this memory is provided by the environment the system interacts with. 
Due to the memory effects in the bath, the time evolution of the system can depend on its state in the past, leading to a non-Markovian dynamics. 
Indeed, in microscopic models including the open system and its environment, the emergence of non-Markovian dynamics is closely related to slowly decaying bath correlation functions, i.e., the fact that the system's influence on the bath persists long enough to impact the future evolution.
However, in particular in master equation approaches, the origin of the memory effects is not as obvious. Moreover, it is known that non-Markovian quantum dynamics can also emerge from classical memory effects and the scenario may not even involve a quantum environment at all~\cite{ CreFac2010, vacchiniClassicalAppraisalQuantum2012,filippovDivisibilityQuantumDynamical2017,MegChrPiiStr2017,megierMemoryEffectsQuantum2021, OppSpaSodWel2023}.
Thus, non-Markovianity in quantum dynamics is not necessarily a genuine quantum memory effect. 
For an arbitrary given quantum dynamics this leads to the question of whether it is actually quantum information that needs to be stored in the environment or if a classical memory would suffice to achieve the same dynamics.
Crucially, standard definitions and witnesses of quantum non-Markovianity, such as non-divisibility or increases in trace distance, merely indicate that a given dynamics is non-Markovian and thus involves some form of memory effect \cite{BreLaiPii2009, RivHuePle2010, HalCreLiAnd2014, BreLaiPiiVac2016}. However, these measures provide no insight into the quantumness of this memory.  

The term \emph{quantum memory} is ambiguous in itself and various incommensurable definitions have been used in the literature~\cite{rossetResourceTheoryQuantum2018,MilEglTarThePleSmiHue2020, GiaCos2021,yuMeasurementDeviceIndependentVerificationQuantum2021,BaeBeyStr2024,TarQuiMurMil2024,Abi2023b, BerGarYadModPol2021,BanMarHorHor2023}.
The definitions mainly differ in their assumptions about what is known to an experimenter who wants to assess the quantumness of the memory in a non-Markovian system. 
If a full description of the global dynamics of system and environment is available, the storage of quantum information in the environmental degrees of freedom could be directly evaluated. 
However, in practice, experimental access to the properties of the environment is usually very limited and the experimenter may try to base their evaluation of the memory's quantumness only on information measurable on the system alone. 

Under this constraint, the most complete picture an experimenter can obtain is given by the so-called process tensor~\cite{PolRodFraPatMod2018, OreCosBru2012, ChiAriPer2009}. This object contains all information about the possible outcomes of arbitrary interventions (for example measurements or quantum channels) the experimenter can perform on the system at predefined times \cite{PolRodFraPatMod2018, MilKimPolMod2019, MilSakPolMod2020, SakTarMil2022, TarEllMil2023, TarPolMilTomMod2019}. 
In particular, the process tensor contains all possible influences of the environment, and it has been shown that the quantumness of the memory is directly related to the process tensor being entangled~\cite{GiaCos2021, TarQuiMurMil2024}.
While the process tensor is in principle accessible by measurements on the system, its experimental determination can be very challenging as it contains all multi-time correlations in the dynamics \cite{GosGiaMonShrRomCos2021}.
Furthermore, the verification of entanglement in the process tensor itself can be difficult for all but the smallest dimensions \cite{GiaCos2021}. 

However, we have shown in Ref.~\cite{BaeBeyStr2024} that the need for quantum memory effects in an environment can be verified even when much less information than the full process tensor is available. 
The criterion in Ref.~\cite{BaeBeyStr2024} is based only on the reduced dynamics in the system, which is not only easier to measure in experiments but makes the approach directly applicable to master equations which also only describe the reduced open system dynamics. 

In this paper, we build on the work in Ref.~\cite{BaeBeyStr2024} and present an entropic witness for the verification of quantum memory in open system dynamics whose great advantage is its practical applicability to systems of arbitrary dimension. 
The original theorem, despite being general in scope, was based on entanglement measures that are often practically intractable for all but the smallest quantum systems like qubits. 
In this work, we present examples beyond the qubit dynamics for which the quantum memory is detectable with the entropic witness.  
In particular, we apply it to Gaussian dynamics of a continuous-variable system, demonstrating that this  criterion is able to detect quantum memory for dynamics in infinite dimensional Hilbert spaces. 

The paper is structured as follows. In the next section, we outline the scenario and define what we mean by classical and quantum memory in this context. 
In Sec.~\ref{sec:old-witness}, we review the quantum memory criterion from Ref.~\cite{BaeBeyStr2024} in a slightly generalized form before we derive the entropic witness in Sec.~\ref{sec:entropic-witness}.
We demonstrate the usefulness of this approach in Sec.~\ref{sec:qudit-example} by means of a qudit example for which the original witness would be hard to compute.
In Sec.~\ref{sec:gaussian-example}, we then turn to the case of Gaussian dynamics, showing that the non-Markovian damping of a harmonic oscillator cannot be realized with classical memory, in general.

\section{Scenario}
\label{sec:scenario}

The central object of interest in our study is the \emph{dynamics} \(\Dyn\) of an open quantum system \(\Sys\).
We define \(\Dyn\) to be a discrete or continuous 
family of completely positive trace-preserving (CPT) maps \(\mathcal{D} = (\cpt_t)\) mapping the system state from the initial time \(t_0\) to time \(t>t_0\).
Assuming that system and environment are uncorrelated at time \(t_0\), this framework covers any physically valid time evolution of \(\Sys\).
Master equation approaches and many microscopic derivations of system-bath dynamics are based on this assumption, and we will stick to it here as well. 

The simplest dynamics that can possibly show memory effects consists of two CPT maps, i.e., \(\Dyn = (\cpt_{t_1},\cpt_{t_2})\). Such a dynamics can be regarded as two snapshots of an underlying time-continuous dynamics.
\begin{defn}
\label{def:classical-memory}
    We call a dynamics \(\Dyn = (\cpt_{t_1},\cpt_{t_2})\) \emph{realizable with classical memory} if there exists a set of Kraus operators \(K_i\), with \(\sum_i K_i^\dagger K_i = \id\) and a set of CPT maps \(\Phi_i\) such that
    \begin{align}
    \label{eq:classical-memory}
        \cpt_{t_1}[\rho] = \sum_i K_i \rho K_i^\dagger, && \cpt_{t_2}[\rho] = \sum_i \Phi_i[K_i \rho K_i^\dagger].
    \end{align}
\end{defn}
This definition has been proposed in Ref.~\cite{BaeBeyStr2024}, a similar construction can be found in Refs.~\cite{LiHalWis2018,TarQuiMurMil2024}.
The reasoning behind the construction is the following. Any CPT map \(\cpt_{t_1}\) can be written in a Kraus representation. Each Kraus operator \(K_i\) can be seen as a measurement operator of a quantum measurement with outcome \(i\).
This outcome \(i\) is classical data and can be stored in a classical memory. The subsequent evolution in the system is given by a quantum channel \(\Phi_i\) that is conditioned on the data $i$ from the classical memory. Being a CPT map, each \(\Phi_i\) can be thought of as being realized with a fresh, uncorrelated environment.

We note that other definitions of quantum memory or similar concepts have been proposed for example in Refs.~\cite{MilEglTarThePleSmiHue2020, GiaCos2021, BaeBeyStr2024, TarQuiMurMil2024, Abi2023b, BerGarYadModPol2021, BusGanGosBadPanMohDasBer2024:p}. They differ from the definition given here primarily in their assumptions on prior knowledge about the environment, for example concerning its dimension. 
The definition used in Ref.~\cite{BaeBeyStr2024} and in this paper is motivated by the question whether a given dynamics is \emph{realizable} by means of classical memory. 
Thus, even if the actual physical environment that causes the dynamics \(\Dyn\) implements the memory in a quantum system rather than a classical one, an experimenter who only has access to the maps \(\cpt_{t_1},\cpt_{t_2}\) must conclude that a quantum memory is \emph{not necessary for the realization} of \(\Dyn\) if it can be written in the form of Def.~\ref{def:classical-memory}.

In general, the question of whether the non-Markovian dynamics can be realized with classical memory, as given in Eq.~\eqref{eq:classical-memory}, is hard to answer, similarly to the separability problem of a mixed quantum state. 
However, we can verify the necessity of a quantum memory by employing suitable witnesses that rule out the existence of a classical realization in the sense of Def.~\ref{def:classical-memory}.
Such a sufficient criterion was proposed in Ref.~\cite{BaeBeyStr2024}.
In the following section, we will review it in a slightly modified form before we continue with the main result of this paper: an entropic witness that is computationally more tractable.

\section{Local witness for quantum memory}
\label{sec:old-witness}
Besides the system of interest \(\Sys\) whose dynamics \(\Dyn\) we would like to analyze, we consider an ancilla quantum system \(\Anc\) that remains unaffected by the dynamics. Initially, the system \(\Sys\) and the ancilla \(\Anc\) are prepared in a joint state \(\rhoSAzero\). 
The system undergoes the dynamics \(\Dyn\) while the ancilla is untouched.
At time \(t\), the joint state of system and ancilla is given by
\begin{align}
\label{eq:choi-state}
    \rhoSA_t = (\cpt_{t}\otimes \id_\Anc)[\rhoSAzero].
\end{align}
Furthermore, we define the following functions 
\begin{align}
\label{eq:F}
    F_f\left[\rhoSA\right] = \min_{\{p_k, \rho_k\}} \sum_k p_k f(\rho_k), \notag\\
    F_f^\sharp\left[\rhoSA\right] = \max_{\{p_k, \rho_k\}} \sum_k p_k f(\rho_k),
\end{align}
where the minimization and maximization run over all decompositions of \(\rhoSA\), i.e., \(\sum_k p_k \rho_k = \rhoSA\), and \(f\) is some non-negative function which is non-increasing under local CPT maps on \(\Sys\).
The necessity of quantum memory for the realization of a dynamics \(\Dyn\) on \(\Sys\) can then be verified by the following theorem.
\begin{theorem}
\label{th:criterion}
    Let \(\cpt_{t_1}\) and \(\cpt_{t_2}\) be two CPT maps on system \(\Sys\) and \(\rhoSAzero\) an initial joint state of \(\Sys\) with an ancilla \(\Anc\). Let \(\rhoSAone\) and \(\rhoSAtwo\) be the joint states at times \(t_1,t_2\) as defined in Eq.~\eqref{eq:choi-state} and \(F_f\) and \(F_f^\sharp\) as in Eq.~\eqref{eq:F}. If we observe
    \begin{align}
    \label{eq:criterion}
        F^\sharp_f\left[\rhoSAone\right] < F_f\left[\rhoSAtwo\right] 
    \end{align}
    for some \(f\), the dynamics \(\Dyn=(\cpt_{t_1},\cpt_{t_2})\) is not realizable with classical memory.
\end{theorem}
In Ref.~\cite{BaeBeyStr2024}, this criterion was proposed for the special case of \(\rhoSAzero\) being initially a maximally entangled state and \(f\) being an entanglement monotone. In this case,  \(\rhoSA_t\) is the Choi state of the channel \(\cpt_t\), \(F\) becomes its entanglement of formation and \(F^\sharp\) its entanglement of assistance~\cite{DiVFucMabSmoThaUhl1999, LauVerEnk2002}.
The proof of Thm.~\ref{th:criterion} is analogous to the one in Ref.~\cite{BaeBeyStr2024} and can be found in App.~\ref{app:proofs}.

The criterion in Eq.~\eqref{eq:criterion} is sufficient but not necessary. It resembles non-Markovianity witnesses based on the entanglement increase with an ancilla~\cite{RivHuePle2010}, the crucial difference being the function \(F^\sharp\) on the left-hand side of Eq.~\eqref{eq:criterion}.
The quantities \(F\) and \(F^\sharp\) are generally hard to compute \cite{Hua2014}. In Ref.~\cite{BaeBeyStr2024}, examples have been given for which \(F\) and \(F^\sharp\) can be calculated if \(f\) is chosen to be the concurrence, but this approach is essentially limited to qubit dynamics.
To overcome this problem of computability, we propose a witness based on entropic bounds that are straightforwardly computable in any dimension.

\section{Entropic witness for quantum memory}
\label{sec:entropic-witness}
The left and the right-hand side of Eq.~\eqref{eq:criterion} can be bounded by entropic quantities if we choose \(f\) to be the entanglement of formation \(E\), defined by 
\begin{align}
    E[\rhoSA] = \min_{\{p_k,\ket{\psi_k}\}} \sum_k p_k S_\Sys(\ketbra{\psi_k}{\psi_k}), 
\end{align}
where \(\sum_k p_k \ketbra{\psi_k}{\psi_k} = \rhoSA\), and \(S_\Sys\) is the von Neumann entropy of the reduced state of system \(\Sys\), i.e, \(S_\Sys(\rhoSA) = -\tr{ \operatorname{tr}_{\Anc}({\rhoSA}) \ln \operatorname{tr}_{\Anc}({\rhoSA})}\). Likewise, \(S_\Anc\) denotes the von Neumann entropy of the reduced state of the ancilla \(\Anc\).
 It was shown in Ref.~\cite{DiVFucMabSmoThaUhl1999} that \footnote{A tighter bound is given by \(F_E^\sharp[\rhoSA] \leq \min\{S_\Sys[\rhoSA],S_\Anc[\rhoSA]\}\). However, this bound does not lead to a tighter inequality~\eqref{eq:entropic-criterion}, so we do not use it here.
}
\begin{align}
\label{eq:eoaupperbound}
    F_E^\sharp\left[\rhoSA\right] \leq S_\Sys\left[\rhoSA\right].
\end{align}
The right-hand side of Eq.~\eqref{eq:criterion} can be bounded by~\cite{CarLie2012}
\begin{align}
	\label{eq:eoflowerbound}
	 F_E\left[\rhoSA\right] &\geq \max\{-S_{\Sys|\Anc}\left[\rhoSA\right], -S_{\Anc|\Sys}\left[\rhoSA\right]\},
\end{align}
where 
\begin{align}
\label{eq:relative-entropy}
    S_{i|j}\left[\rhoSA\right] :=S\left[\rhoSA\right] - S_j\left[\rhoSA\right]
\end{align}
is the conditional quantum entropy, and \(S[\rhoSA]\) denotes the von Neumann entropy of the joint state.
Invoking Eqns.~\eqref{eq:criterion},\eqref{eq:eoaupperbound}, and \eqref{eq:eoflowerbound}, we arrive at an inequality for witnessing quantum memory.
\begin{theorem}
	\label{th:entropictheorem}
	Let $\cpt_{t_1}$ and $\cpt_{t_2}$ be two CPT maps on system \(\Sys\) and \(\rhoSAzero\) an initial joint state of \(\Sys\) with an ancilla \(\Anc\). Let \(\rhoSAone\) and \(\rhoSAtwo\) be the joint states at times \(t_1,t_2\) as defined in Eq.~\eqref{eq:choi-state}. If we observe
	\begin{align}
 \label{eq:entropic-criterion}
	S_\Sys\!\left[\rhoSAone\right] < \max\left\{-S_{\Sys|\Anc}\!\left[\rhoSAtwo\right], -S_{\Anc|\Sys}\!\left[\rhoSAtwo\right]\right\}
\end{align}
    the dynamics $\Dyn = \dynamics{\cpt_{t_1}, \cpt_{t_2}}$ is not realizable with classical memory.
\end{theorem}

These entropic quantities can easily be computed for finite quantum systems. We will give an example in the next section. 
We then extend the approach to the infinite dimensional case and consider  Gaussian dynamics whose quantum memory can be verified with the entropic criterion, too.

\section{Example I: Qudit dynamics}
\label{sec:qudit-example}
As a first example, we apply the quantum memory criterion to a class of dynamics describing non-Markovian damping of a qudit system \(\Sys\) of dimension \(d\)~\cite{reichExploitingNonMarkovianityQuantum2015}.
The system is coupled to a single memory qubit \(\Mem\), which in turn is damped by a Markovian bath at zero temperature. Tracing out the memory qubit, we are interested in the extent to which the entropic criterion \eqref{eq:entropic-criterion} applied to the local dynamics of \(\Sys\) alone can detect the quantumness of the memory in the environment. 
For \(d=2\), this model reduces to the well-studied non-Markovian qubit amplitude damping master equation~\cite{breuerTheoryOpenQuantum2007, KreLuoStr2016, diosiNonMarkovianQuantumState1998, garrawayNonperturbativeDecayAtomic1997}, which has been shown to require quantum memory~\cite{BaeBeyStr2024}. 

\begin{figure}[ht]
    \centering
    \includegraphics[width=\linewidth]{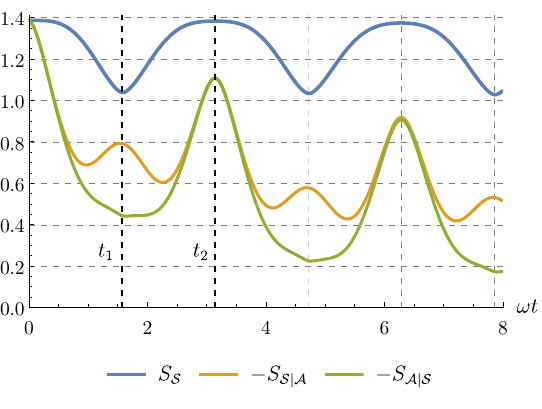}
    \caption{Time evolution of the entropic quantities used in Eq.~\eqref{eq:entropic-criterion} for a system of dimension \(d=4\) and \(\gamma/\omega = 0.05\). The initial state \(\rhoSAzero\) of system and ancilla in Eq.~\eqref{eq:choi-state} is chosen to be the maximally entangled state \(\ket{\Phi_+} = \sum_{l=0}^{d-1} \ket{ll}/\sqrt{d}\). The minimum of \(S_\Sys\) at time \(t_1\) is smaller than the subsequent maximum of \(-S_{S|A}\) and \(-S_{A|S}\) at time \(t_2\). Thus, quantum memory is demonstrated by the entropic criterion in Eq.~\eqref{eq:entropic-criterion}. The Markovian damping of the memory qubit [see Eq.~\eqref{eq:GKSL}] prevents us from revealing the quantum memory at later times. The second revival of the conditional entropies is already too small to satisfy Eq.~\eqref{eq:entropic-criterion}.}
    \label{fig:time-evolution}
\end{figure}

The interaction Hamiltonian between the system \(\Scal\) and the memory \(\Mem\) is given by 
\begin{align}
\label{eq:GlobalUnitary}
    H_{\Sys\Mem} = \omega \left(J_-^{(d)} \otimes \sigma_+ + J_+^{(d)} \otimes \sigma_-\right),
\end{align}
 where $J_\pm^{(d)}$ are the $d$-dimensional ladder operators and $\sigma_\pm=J_\pm^{(2)}$.
The damping of \(\Mem\) is given by a single Lindblad dissipator with constant rate \(\gamma\). Thus, the Gorini–Kossakowski–Sudarshan–Lindblad (GKSL) equation for the joint state of system and memory is given by
\begin{align}
\label{eq:GKSL}
    \dot\rho_{\Sys\Mem}  &= -\i [H_{\Sys\Mem}, \rho_{\Sys\Mem}] + \gamma \Dcal[\id_\Sys \otimes \sigma_-]\rho_{\Sys\Mem}\notag\\
    &= \Lcal \rho_{\Sys\Mem} ,
\end{align}
where \(\Dcal[X]\rho = X\rho X^\dagger - (X^\dagger X \rho + \rho X^\dagger X)/2\).
The dynamics \(\cpt_t\) of the system \(\Sys\) is obtained by integrating this master equation for an initial memory state \(\rho_\Mem = \ketbra{0}{0}\) up to time \(t\) and tracing over \(\Mem\)
\begin{align}
    \cpt_t[\rho_\Sys] = \ptr{\Mem}{e^{\Lcal t}\rho_\Sys\otimes\rho_\Mem}.
\end{align}
For \(d>2\), the system dynamics \(\cpt_t\) does not have a simple closed form. The following results were obtained numerically.
In Fig.~\ref{fig:time-evolution}, we plot the entropic quantities involved in inequality~\eqref{eq:entropic-criterion} over time for a system of dimension \(d=4\) and a ratio between damping and coupling of \(\gamma/\omega = 0.05\). 
The ancilla \(\Anc\) has the same dimension as the system and the initial system-ancilla state is chosen to be the maximally entangled state \(\rhoSAzero = \ketbra{\Phi_+}{\Phi_+}\), with \(\ket{\Phi_+} = \sum_{l=0}^{d-1} \ket{ll}/\sqrt{d}\).
 
The optimal times \(t_1\) and \(t_2\) for the detection of quantum memory in this particular example are marked. 
At time \(t_2\), the right-hand side of Eq.~\eqref{eq:entropic-criterion} is greater than the left-hand side at time \(t_1\). The second revival in the plot is already too small to witness quantum memory with the entropic inequality~\eqref{eq:entropic-criterion}. 

\begin{figure}[ht]
    \centering
    \includegraphics[width=1\linewidth]{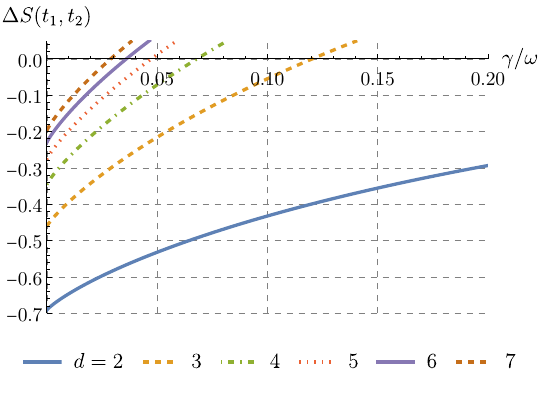}
    \caption{We plot \(\Delta S\) from Eq.~\eqref{eq:deltaS} as a function of \(\gamma/\omega\) for different dimensions \(d\) of the qudit system \(\Sys\). The times \(t_1\) and \(t_2\) in Eq.~\eqref{eq:deltaS} are chosen to minimize \(\Delta S\) for the respective parameters (see main text and Fig.~\ref{fig:time-evolution}). 
    For \(d>2\) there is a critical \(\gamma/\omega\) above which the verification of the quantum memory by means of the entropic witness becomes impossible for the given dynamics.
    }
    \label{fig:qudit_damping}
\end{figure}

Generally, the satisfiability of Eq.~\eqref{eq:entropic-criterion} sensitively depends on the ratio \(\omega/\gamma\). A stronger damping washes out the signature and quantum memory becomes undetectable. 
For the maximally entangled initial state chosen here, we always have \(-S_{\Sys|\Anc}\!\left[\rhoSAtwo\right] \geq -S_{\Anc|\Sys}\!\left[\rhoSAtwo\right]\).
Thus, by using Eq.~\eqref{eq:entropic-criterion} and defining the witness
\begin{align}
\label{eq:deltaS}
    &\Delta S(t_1,t_2) = S_\Sys\!\left[\rhoSAone\right] +S_{\Sys|\Anc}\!\left[\rhoSAtwo\right],
\end{align}
quantum memory is demonstrated for \(\Delta S < 0\).
In Fig.~\ref{fig:qudit_damping}, we plot the witness \(\Delta S(t_1,t_2)\) as a function of the damping-to-coupling ratio \(\gamma/\omega\) for different dimensions \(d\). Similarly to Fig.~\ref{fig:time-evolution}, \(t_1\) is always chosen to be the first minimum of \(S_\Sys\) and \(t_2\) the first subsequent maximum of \(-S_{\Sys|\Anc}\) for the given \(d\) and \(\gamma/\omega\) .
We see that the critical \(\gamma/\omega\), below which quantum memory can be witnessed, decreases with increasing dimension. Only in the qubit case \(d=2\) can the quantumness of the memory be demonstrated for arbitrary finite \(\gamma/\omega\). 

Due to the inequalities in Eq.~\eqref{eq:eoaupperbound} and \eqref{eq:eoflowerbound} the entropic witness is in general looser than the entanglement witness in Ref.~\cite{BaeBeyStr2024}. 
However, it must be emphasized that the entropic witness is able to verify the quantumness of the memory for system dimensions for which the entanglement quantities in the witness in Ref.~\cite{BaeBeyStr2024} could hardly be calculated efficiently, making it practically applicable to systems beyond the qubit.

\section{Gaussian dynamics}
\label{sec:gaussian-example}
The entropic quantities used in Thm.~\ref{th:entropictheorem}
are well-defined also for an infinite (separable) Hilbert space \cite{CarLie2012, LieRus1973, Shi2016}. Thus,
this form of the quantum memory criterion can also be applied to continuous-variable (CV) systems. We demonstrate this power here using non-Markovian Gaussian channels.
Non-Markovian dynamics in CV systems have been studied extensively, e.g.\ in \cite{VasOliParMan2011,Abi2023b, VasManParBrePii2011, heContinuousvariableQuantumTeleportation2011,torreNonMarkovianityGaussianChannels2015,souzaGaussianInterferometricPower2015,groblacherObservationNonMarkovianMicromechanical2015,liuzzo-scorpoNonMarkovianityHierarchyGaussian2017,torreExactNonMarkovianDynamics2018, RicWieBre2022}. 
In the following, we show how the necessity of quantum memory for the realization of such a dynamics can be witnessed by means of Eq.~\eqref{eq:entropic-criterion} and provide illustrative examples.

Let us first review relevant properties of Gaussian quantum states, following the notation in~\cite{SerIllSie2003}. We consider an $N$-mode CV system with Hilbert space $\Hilb = \bigotimes_{i=1}^{N} \Hilb_i$. Each infinite dimensional Hilbert space $\Hilb_i$ is equipped with creation and annihilation operators obeying the usual bosonic commutation relations $[\op{a}_i, \op{a}_j^\dagger]=\delta_{ij}$, and $[\op{a}_i, \op{a}_j]=[\op{a}_i^\dagger, \op{a}_j^\dagger]=0$. It is common to introduce position and momentum quadratures
\begin{align}\label{eq:qp}
	\op{q}_i = \frac{1}{\sqrt{2}} (\op{a}_i+\op{a}_i^\dagger), \qquad 	\op{p}_i = \frac{\i}{\sqrt{2}} (\op{a}_i^\dagger-\op{a}_i)
\end{align}
and combine them to a $2N$-dimensional phase space vector $\op{\vec X}=(\op{q}_1, \op{p}_1, \dots, \op{q}_N, \op{p}_N)$ such that the canonical commutation relations take the form
\begin{align}
	\comm{\op{X}_i, \op{X}_j} = \i \Omega_{ij},\;\;i,j = 1,\ldots, 2N,\;\; && \Omega = \bigoplus_{i=1}^{N}
	\begin{pmatrix}
		0 & 1\\ -1 &0
	\end{pmatrix} .
\end{align}
Gaussian quantum states $\rho$ are 
fully determined by their position-momentum mean value $\langle\vec{X}\rangle = \tr{\op{\vec{X}}\rho}$ and the corresponding (symmetrized) covariance matrix
\begin{align}\label{eq:covmat}
	\covmat_{ij} = \frac{1}{2} \langle \op{X}_i \op{X}_j + \op{X}_j \op{X}_i \rangle -\langle \op{X}_i \rangle \langle \op{X}_j \rangle.
\end{align}
{As a shift of the mean value amounts to a local unitary transformation that leaves entropic quantities invariant, we may assume, without loss of generality, that it is zero in the following, $\langle\vec{X}\rangle=0$.}
A valid density matrix satisfies $\covmat + \i \Omega/2 \geq 0$.

A Gaussian channel ${\cal E}: \rho\rightarrow \rho'= {\cal E}[\rho]$ maps Gaussian quantum states onto themselves and can thus be expressed as a linear map of covariance matrices
that takes the form~\cite{SerIllSie2003, OskManWin2022}
\begin{align}\label{eq:gausschannel}
	\covmat \mapsto \covmat' = M^\top \covmat M + N.
\end{align}
Here, the matrices $M$ and $N$ have to satisfy
\begin{align}
	N + \frac{\i}{2} \Omega - \frac{\i}{2} M^\top \Omega M \geq 0
\end{align}
in order to describe a completely positive Gaussian channel~\cite{OskManWin2022}.

We can now adapt the inequality (\ref{eq:entropic-criterion}) for Gaussian dynamics \(\Dyn\) of a system \(\Sys\), i.e., a family of Gaussian quantum channels \(\Ecal_t\) described by matrices \(M_t,N_t\). 
We need an initial joint Gaussian state \(\rhoSAzero\) between the system \(\Sys\) and an ancilla \(\Anc\) [see Eq.~\eqref{eq:choi-state}].
Its covariance matrix can be written as
\begin{align}
    \sigmaSA_0 = \begin{pmatrix}
        \bm\alpha_0 && \bm\gamma_0 \\
        \bm\gamma_0^\top && \bm\beta_0
    \end{pmatrix},
\end{align}
where \(\bm\alpha_0\) and \(\bm\beta_0\) are the covariance matrices of the reduced states of \(\Sys\) and \(\Anc\), respectively, and \(\bm\gamma_0\) describes the correlations between the two.
Applying the Gaussian channel \(\Ecal_t\) to the subsystem \(\Sys\) we obtain a joint state \(\rhoSA_t\) whose covariance matrix reads 
\begin{align}\label{eq:choi_covariance}
    \sigmaSA_t \!=\! \begin{pmatrix}
        \bm\alpha_t && \bm\gamma_t \\
        \bm\gamma^\top_t && \bm\beta_t
    \end{pmatrix} \!=\! \begin{pmatrix}
        M_t^\top \bm\alpha_0 M_t + N_t && M_t^\top \bm\gamma_0 \\
        \bm\gamma_0^\top M_t && \bm \beta_0
    \end{pmatrix}.
\end{align}
The entropies of the joint and the reduced states in Eq.~\eqref{eq:entropic-criterion} of Thm.~\ref{th:entropictheorem} can then be calculated directly from the symplectic eigenvalues of the matrices \(\sigmaSA_t, \bm\alpha_t\), and \(\bm\beta_t\) at times \(t_1\) and \(t_2\) (see Refs.~\cite{weedbrookGaussianQuantumInformation2012,holevoCapacityQuantumGaussian1999} for details) and with Eq.~\eqref{eq:relative-entropy}, the criterion for quantum memory reads
\begin{align}
\label{eq:DeltaS-gaussian}
    \Delta S(t_1,t_2) = &S[\bm\alpha_{t_1}] + S[\sigmaSA_{t_2}] \notag \\
      -&\max\{S[\bm\alpha_{t_2}],S[\bm\beta_{t_2}]\} < 0.
\end{align}

\section{Example II: Single-mode dynamics}
Let us focus on the special case of both system \(\Sys\) and ancilla \(\Anc\) being single modes for which the quantum memory criterion becomes particularly simple.
The von Neuman entropy of the single-mode Gaussian system state with covariance matrix \(\bm\alpha\) (and analogously for the reduced ancilla state described by \(\bm\beta\)) is given by~\cite{SerIllSie2003}
\begin{align}
    S_1(\bm\alpha) = h\left(\sqrt{\det \bm\alpha}\right),
\end{align}
with
\begin{align}\label{eq:ffunction}
	h(x) = \left(x+ \frac12\right) \ln\left(x + \frac12\right) - \left(x - \frac{1}{2}\right)\ln\left(x-\frac{1}{2}\right).
\end{align}
The entropy of the joint two-mode Gaussian state reads~\cite{SerIllSie2003}
\begin{align}
	\label{eq:s2modevn}
	S_2(\sigmaSA) = h[n_-(\sigmaSA)]+ h[n_+(\sigmaSA)],
\end{align}
with
\begin{align}
	n_\mp(\covmat) = \sqrt{\frac{\Delta(\covmat) \mp \sqrt{\Delta(\covmat)^2 - 4 \det(\covmat)}}{2}},
\end{align}
and \(\Delta(\covmat) = \det(\bm\alpha) + \det(\bm\beta) + 2 \det(\bm\gamma)\).

For the sake of concreteness, let us consider an initial two-mode squeezed state of system and ancilla whose covariance matrix is of the form
\begin{align}
\label{eq:sigma-S-A}
    \sigmaSA_0 = \frac{1}{2}\begin{pmatrix}
        A_r && B_r \\
        B_r && A_r
    \end{pmatrix},
\end{align}
with
\begin{align}
    A_r = \cosh r \id, && B_r = \sinh r \sigma_z,
\end{align}
and $r>0$ the squeezing parameter. 
The CV equivalent of a
maximally entangled Bell state is obtained in the limit \(r \to \infty\).
The time-evolved state is then described by
\begin{align}
\label{eq:abc-squeezed}
    \bm\alpha_t  = \frac{1}{2}M_t^\top A_r M_t +  N_t, && \bm\beta_t = \frac{A_r}{2}, && \bm\gamma_t = \frac{M_t^\top B_r}{2}.
\end{align}
In the following we evaluate the quantum memory witness for  Gaussian dynamics describing energy loss of a single mode. 

\subsection{Lossy Gaussian Channels}
\label{sec:lossy}
\begin{figure}
    \centering
    \includegraphics[width=1\linewidth]{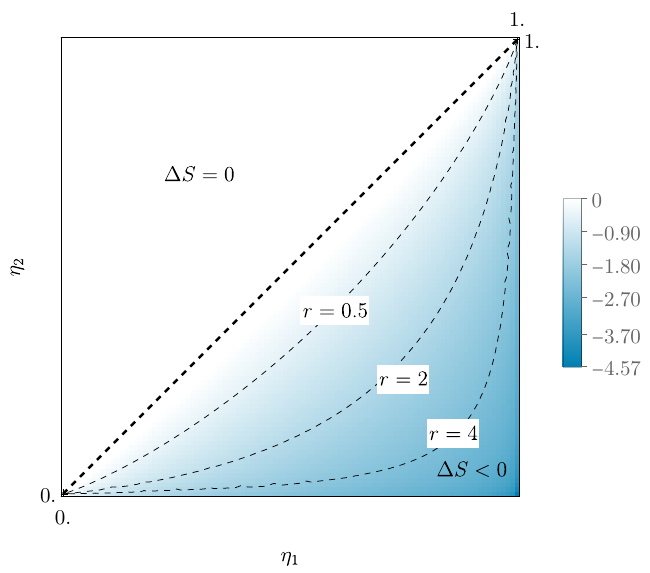}
    \caption{Minimal value of $\Delta S$ obtained by varying $r>0$ for every combination of $\eta_1$ and $\eta_2$. The diagonal dashed line describes the case that $\eta_1 = \eta_2$ and every combination of $\eta_1$ and $\eta_2$ lying beneath this line correspond to dynamical maps $\cpt_{t_1}$ and $\cpt_{t_2}$ which can only be realized using quantum memory since there is an arbitrary small $r>0$ such that $\Delta S<0$. In general, the detection of quantum memory gets easier for larger differences of \(\eta_1\) and \(\eta_2\). 
    We also plot boundaries \(\Delta S = 0\) for three different fixed values of the squeezing parameter \(r\) (dashed lines). Stronger squeezing shrinks the parameter region for which quantum memory can be detected. 
    This is interesting since it tells us that greater entanglement in the initial state \(\rhoSAzero\) diminishes the ability of the entropic criterion to witness quantum memory.}
    \label{fig:lossychannel}
\end{figure}

A simple dynamics modelling damping in a Gaussian mode is given by
\begin{align}
\label{eq:lossy-channel}
   \cpt_{t}: \quad M_{t}= \sqrt{1-\eta_t} \id \qquad  N_{t} = \eta_t \frac{\id}{2},
\end{align}
where \(0\leq \eta \leq 1\) is the loss parameter \cite{EisWol2007}. Full loss is reached for $\eta=1$, whereas $\eta=0$ is the identity channel. We do not further specify the time evolution of \(\eta_t\) at this point, but only assume that the losses at time \(t_1\) and \(t_2\) are described by \(\eta_1\) and \(\eta_2\), respectively.
For \(\eta_2 \geq \eta_1\), i.e., a situation where the loss grows over time, the dynamics is Markovian and therefore does not require memory at all. 
For the opposite case of a partially reversed loss \(\eta_2 < \eta_1\) the dynamics is not only non-Markovian but also requires quantum memory as can be shown with our entropic witness. 
Plugging Eq.~\eqref{eq:lossy-channel} into Eq.~\eqref{eq:abc-squeezed}, we find for \(\Delta S\) in Eq.~\eqref{eq:DeltaS-gaussian}
\begin{align}
    \Delta S&(\eta_1,\eta_2)\notag = h\!\left(\frac{\eta_1 + (1-\eta_1)\cosh r}{2}  \right) +   \notag \notag\\* &+ h\!\left(\frac{1-\eta_2 + \eta_2 \cosh r}{2} \right) - h\!\left(\frac{\cosh r}{2} \right). 
\end{align}
The ability to demonstrate the quantumness of the memory depends on the squeezing parameter \(r\) of the initial system-ancilla state in Eq.~\eqref{eq:sigma-S-A}. 
We plot \(\Delta S\) in Fig.~\ref{fig:lossychannel}. For each pair~\((\eta_1, \eta_2)\), \(\Delta S\) was minimized over the squeezing parameter \(r\) and we see that quantum memory can indeed be witnessed for any choice \(\eta_2< \eta_1\). Intuitively, the detection of quantum memory becomes easier for greater differences between the loss parameters \(\eta_1\) and \(\eta_2\). 

We also plot the boundary \(\Delta S = 0\) for several fixed choices of \(r\). Interestingly, stronger squeezing, and therefore stronger initial entanglement between \(\Sys\) and \(\Anc\), decreases the parameter region for which the entropic witness can verify quantum memory.  

\subsection{Non-Markovian damped harmonic oscillator}
Next, let us take a look at a common, time-continuous open quantum system dynamics that essentially leads to the lossy channel described above:
A damped harmonic oscillator (operators $a, a^\dagger$), whose dynamics is given by the master equation
\begin{align}
    \label{eq:mastereq}
	\dot{\rho} = - \i \comm{\omega_t a^\dagger a, \rho} +  \frac{\gamma_t}{2}\left( \comm{a, \rho a^\dagger} + \comm{a \rho, a^\dagger}\right).
\end{align}
For constant positive parameters $\omega_t=\text{const.}>0$ and $\gamma_t=\text{const.}>0$, this is the well known GKSL generator for a Markovian semigroup dynamics. If, however, the damping rate is negative during certain time intervals ($\gamma_t<0$), the master equation describes non-Markovian dynamics~\cite{torreNonMarkovianityGaussianChannels2015}. In order to generate a proper CPT dynamical map, restrictions on the function $\gamma_t$ apply. Below, we will construct admissible functions $\omega_t$ and $\gamma_t$ from an explicit microscopic model.

The generator being quadratic in annihilation and creation operators, we conclude that the master equation (\ref{eq:mastereq}) induces a Gaussian dynamical map. For the second-order moments $\langle a^\dagger a\rangle_t$, $\langle a a\rangle_t$, and $\langle a^\dagger a^\dagger\rangle_t$ it is straightforward to determine the Heisenberg equations of motion from Eq.~(\ref{eq:mastereq}). Their solutions turn out to be simple time-dependent exponentials with
\begin{align}\label{eq:timeevolution}
    \langle a^\dagger a\rangle_t & =  \langle a^\dagger a\rangle_0 e^{-\Gamma_t},\\ \nonumber
    \langle a a\rangle_t & =  \langle a a\rangle_0 e^{-2\i\Phi_t-\Gamma_t},\\ \nonumber
       \langle a^\dagger a^\dagger\rangle_t & =  \langle a^\dagger a^\dagger\rangle_0 e^{+2\i\Phi_t-\Gamma_t}.
\end{align}
Here, the integrated phase and damping terms follow naturally from the parameters of the master equation and read
\begin{align}
\Phi_t  =  \int_0^t \d s\, \omega_s,\;\;
    \Gamma_t  = \int_0^t \d s\, \gamma_s \, .
\end{align}

It is now straightforward to relate the covariance matrix $\covmat_t$ in Eq.~(\ref{eq:covmat}) expressed in terms of $(q_t,p_t)$ to its initial matrix $\covmat_0$ making use of the usual relations in Eq.~(\ref{eq:qp}) and the results in Eq.~(\ref{eq:timeevolution}). We find the Gaussian channel relation $\covmat_t = M_t^\top \covmat_0 M_t + N_t$ from Eq.~(\ref{eq:gausschannel}) with~\cite{torreNonMarkovianityGaussianChannels2015}
\begin{align}
	\label{eq:MmatDampedHO}
	M_t&= e^{-\Gamma_t/2} \begin{pmatrix}
		\cos \Phi_t & -\sin \Phi_t\\
		\sin \Phi_t & \phantom{-}\cos \Phi_t
	\end{pmatrix},\\ \nonumber
	N_t &= \left( 1 - e^{-\Gamma_t}\right) \frac{\id}{2}.
\end{align}
The time dependent phase $\Phi_t$ appearing in the rotation matrix of $M_t$ reflects the Hamiltonian contribution to the dynamics in Eq.~\eqref{eq:mastereq}. Comparing Eq.~\eqref{eq:MmatDampedHO} to Eq.~\eqref{eq:lossy-channel} we see that the damped harmonic oscillator model describes a time-continuous lossy channel with
\begin{align}
    \eta_t =1 - e^{-\Gamma_t},
\end{align}
that undergoes an additional unitary rotation in phase space described by $\Phi_t$.
Since the entropies in Eq.~\eqref{eq:DeltaS-gaussian} are invariant under local unitaries on the system, the quantumness of the memory involved in the dynamics is solely determined by the function \(\Gamma_t\). We know from Fig.~\ref{fig:lossychannel} that whenever $\eta_t$ is a non-monotonic function, the corresponding dynamics requires quantum memory.

Let us now specify the non-Markovian Gaussian channel by fixing the functions $\omega_t$ and $\gamma_t$ of its master equation. In order to ensure a CPT dynamical map, we determine those functions from the full, unitary system-environment dynamics of an oscillator (operators $a,a^\dagger$) coupled to a
bath of such oscillators (operators $b_\lambda, b_\lambda^\dagger$) in their ground state (zero temperature). The mode $a$ could be a cavity mode, then the $b_\lambda$-modes would correspond to the environmental electromagnetic modes, in their ground state. The total system-environment Hamiltonian is chosen to be
\begin{align}\label{eq:totalH}
    H_\text{tot} = \omega a^\dagger a + \sum_\lambda (g_\lambda  ab_\lambda ^\dagger + g_\lambda^*  a^\dagger b_\lambda) + \sum_\lambda \omega_\lambda b_\lambda^\dagger b_\lambda.
\end{align}
For this model, the reduced dynamics of the system oscillator follows an exact master equation~\cite{PurLaw1977},
\begin{align}
    \label{eq:mastereq0}
	\dot{\rho} = - \i \comm{\omega a^\dagger a, \rho} +  G^*_t \comm{a, \rho a^\dagger} +  G_t\comm{a \rho, a^\dagger}.
\end{align}
Crucially, setting
\begin{align}
    \omega_t := \omega + \operatorname{Im}(G_t),\;\;\;\gamma_t := 2 \operatorname{Re}(G_t),
\end{align}
the exact master equation \eqref{eq:mastereq0} takes the desired form of Eq.~\eqref{eq:mastereq}.
Here, the complex, time-dependent coefficient $G_t$ can be written in the form
\begin{align}\label{eq:Gfunction}
	G_t = \frac{\int_{0}^{t} \d s\; \alpha(t-s) c_s}{c_t},
\end{align}
where
\begin{align}
    \alpha(t-s) = \sum_\lambda |g_\lambda|^2 e^{-\i\omega_\lambda(t-s)}
\end{align}
is the (zero temperature) bath correlation function of the underlying system-bath model \eqref{eq:totalH}. The
complex amplitude $c_t$ satisfies the linear equation of motion
\begin{align}\label{eq:amplitude}
	\dot{c}_t = -\i \omega c_t - \int_{0}^{t} \d s\, \alpha(t-s)\, c_s.
\end{align}
In what follows, we set the initial condition to $c(0)=1$ ($G_t$ and therefore $\rho_t$ is in fact independent of the initial condition).
It can then be seen that
\begin{align}
|c_t|^2 = e^{-\Gamma_t}, 
\end{align}
providing a direct way to determine the channel parameter $\Gamma_t$ directly from the evolution equation \eqref{eq:amplitude} for the amplitude $c_t$.
Depending on the choice of the bath correlation function, the rate $\gamma_t$ may indeed become negative at times and thus the model can describe non-Markovian quantum dynamics.

\begin{figure}
    \centering
    \includegraphics[width=\linewidth]{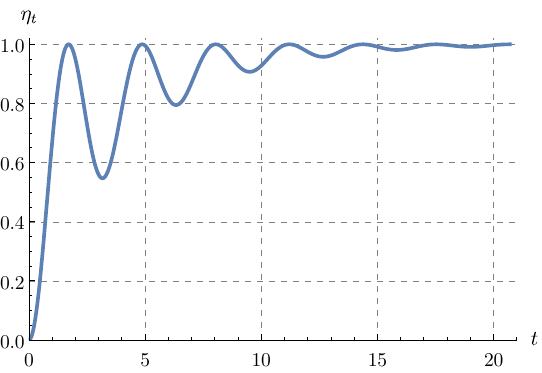}
    \caption{The loss parameter \(\eta_t = 1 - \abs{c_t}^2\) for $|g|^2=1$, $\kappa=1/4$, $\omega=1$, and $\Omega=1$.  The non-monotonous behavior cannot be explained by classical memory effects, as we can always find two times \(t_2 > t_1\) for which \(\eta_{t_2} < \eta_{t_1}\), a signature of quantum memory as shown in Sec.~\ref{sec:lossy}.}
    \label{fig:c_dampedho}
\end{figure}

We now fix our model by specifying the bath correlation function. We choose the exponentially decaying
\begin{align}\label{eq:ouprocess}
	\alpha(t-s) = |g|^2 \e^{-\kappa (t-s) - \i \Omega(t-s)},
\end{align}
with constant parameters \(\kappa, \Omega\).
Then, with Eq.~\eqref{eq:amplitude}, the relevant amplitude $c_t$ follows from the time-local equation
\begin{align}
	\ddot{c}_t + (\kappa + \i \omega + \i \Omega) \dot{c}_t + [|g|^2 + \i \omega (\kappa + \i \Omega)] c_t = 0
\end{align}
with the initial conditions $c_0=1$ and $\dot{c}_0 = - \i \omega$.
Solving this ODE, we can display $\eta_t = 1 - e^{-\Gamma_t} = 1 - |c_t|^2$ as depicted in Fig.~\ref{fig:c_dampedho}.

In Sec.~\ref{sec:lossy} we saw that two consecutive lossy channels \(\cpt_{t_1}\) and \(\cpt_{t_2}\) with $\eta_2 < \eta_1$
cannot be connected via classical memory.
Therefore, noting that $\eta_t$ is a non-monotonous function for our choice of the bath correlation function, see Fig.~\ref{fig:c_dampedho}, we can conclude that the master equation~\eqref{eq:mastereq} based on the choice of bath correlation function \eqref{eq:ouprocess} does indeed describe a non-Markovian evolution where quantum memory is fundamentally involved: there is no way to arrive at the general form of the non-Markovian master equation \eqref{eq:mastereq} from a model involving only classical memory. This is in contrast, for instance, to the master equation of eternal non-Markovianity, that, by construction, may be obtained from random unitary dynamics or the probabilistic mixing of Markovian evolutions, see~\cite{MegChrPiiStr2017}, and thus need not involve any quantum (environmental) memory.

\section{Conclusion}
\label{sec:conclusion}
Non-Markovian quantum dynamics can arise from the presence of either a classical or a quantum memory. A key question in this context is how to distinguish the nature of the memory using only local information about the dynamics. Existing criteria often rely on entanglement measures, which are notoriously challenging to compute for systems beyond qubits.
In this article we propose a tractable way to witness quantum memory in quantum systems extending beyond qubits. 
We start from the definition of quantum memory provided in Ref.~\cite{BaeBeyStr2024} and generalize a sufficient criterion of quantum memory presented there.
Based on this witness, an entropic criterion to verify the quantumness of the memory of the dynamics is derived. This criterion does not only hold in any dimension but is easily computable.

We provide several examples of its application.
First, the entropic witness is used to show that non-Markovian damping dynamics of a qudit requires quantum memory, in general.
In order to illustrate that the entropic witness can also be used for dynamics of infinite-dimensional systems, we show how to witness the quantumness of the memory in Gaussian processes. 
As an example, a model of a lossy channel is investigated. A time-continuous realization of this dynamics is given by a damped harmonic oscillator. By considering the corresponding master equation, we show that the memory involved in this process is of quantum nature.

The entropic witness is a sufficient but not a necessary criterion for quantum memory. 
While it is derived as a bound of the original criterion in Ref.~\cite{BaeBeyStr2024} and is, thus, in principle, less sensitive to quantum memory than the original, we demonstrate in this contribution that it remains capable of detecting quantum memory across a broad class of open system dynamics. Moreover, its computational simplicity makes it a practical and effective tool for characterizing the nature of the memory in non-Markovian quantum dynamics.

\bibliography{literature_memory_gauss}

\appendix

\section{Proof of Theorem~\ref{th:criterion}}
\label{app:proofs}
Suppose we are given a dynamics $\Dyn=\dynamics{\cpt_{t_1}, \cpt_{t_2}}$ on the system \(\Sys\) which requires only classical memory. 
A joint state  $\rhoSAzero$ of the system  with an otherwise untouched ancilla \(\Anc\), evolves to a state \(\rhoSA_{t}\) as given in Eq.~\eqref{eq:choi-state}.
According to Def.~\ref{def:classical-memory}, at time \(t_1\) there is a decomposition $\{p_i, \rho_i\}$ of this state \(\rhoSAone\) given by
\begin{align}
\label{eq:decomposition-t1}
    \rho_i= \frac{1}{p_i}(K_i \otimes \id_\Anc )\rhoSAzero (K_i^\dagger \otimes \id_\Anc )
\end{align}
where $\{K_i\}$ is a local measurement implementing the first map $\cpt_{t_1}$ on the system, \(p_i = \tr{(K_i^\dagger K_i \otimes \id_\Anc)\rhoSAzero}\), and \(\sum_i p_i \rho_i = \rhoSAone\).
Using the quantities defined in Eq.~\eqref{eq:F} we can write
\begin{align}
    F_f^\sharp[\rhoSAone] = \max_{\{p_k, \rho_k\}} \sum_k p_k f(\rho_k) \geq \sum_i p_i f(\rho_i),
\end{align}
where \(\sum_k p_k \rho_k = \rhoSAone\), and, by definition of the maximum, the decomposition labelled by \(i\) and given in Eq.~\eqref{eq:decomposition-t1} leads to a smaller value.
We now define 
\begin{align}
    \rho_i' = \left(\Phi_i \otimes \id_\Anc\right)[\rho_i].
\end{align}
According to Def.~\ref{def:classical-memory}, the \(\rho_i'\) decompose the system-ancilla state at time \(t_2\), i.e.,
\begin{align}
    \sum_i p_i \rho_i' =  \rhoSAtwo.
\end{align}
Since $f$ is a function which is non-increasing under the local quantum channels \(\Phi_i\), we have
\begin{align}
    \label{eq:proof_bound}
    \sum_i p_i f(\rho_i) \geq \sum_i p_i f(\rho_i'),
\end{align}
The right-hand side in Eq.~\eqref{eq:proof_bound} can further be lower bounded. By the definition of the minimum we have
\begin{align}
    \sum_{i}p_i f(\rho_i') &\geq \min_{\{p_l, \rho_l\}} \sum_{l} p_l f(\rho_l)
	=F_f\left[\rhoSAtwo\right],
\end{align}
where \(\sum_l p_l \rho_l = \rhoSAtwo\).

Thus, under the assumption of classical memory given in Def.~\ref{def:classical-memory}, we obtain the inequality
\begin{align}
    F_f^\sharp[\rhoSAone] \geq F_f\left[\rhoSAtwo\right].
\end{align}
Any violation of this inequality therefore demonstrates the necessity of quantum memory. This is Theorem~\ref{th:criterion}. \hfill \(\blacksquare\)

\end{document}